\documentclass[a4paper]{jpconf}
\usepackage{graphicx}

\usepackage{color}

\begin{document}
\title{Stellar evolution in the outer Galaxy}

\author{Ryszard Szczerba$^1$, Natasza Si\'odmiak$^1$, Aleksandra Le\'sniewska$^2$, Agata Karska$^2$ and Marta Sewi\l o$^{3,4}$}

\address{$^1$ N. Copernicus Astronomical Center, PAS, Rabia\'nska 8, 87-100 Toru\'n, Poland \\
$^2$ Astronomical Observatory, Adam Mickiewicz University, S{\l}oneczna 36, 60-268 Pozna\'n, Poland\\
$^3$ NASA Goddard Space Flight Center, 8800 Greenbelt Rd., Greenbelt, MD 20771, USA\\
$^4$ Astronomical Observatory of the Jagiellonian University, Orla 171, 30-244 Krak\'ow, Poland
}

\ead{$^1$ szczerba@ncac.torun.pl}

\begin{abstract}
We investigate the distribution of different classes of spectroscopically identified sources and theoretical models in the color-color diagrams (CCDs) combining the near-infrared (NIR) and mid-infrared (MIR) data to develop a method to classify Outer Galaxy sources detected with the \emph{Spitzer} Space Telescope (hereafter \emph{Spitzer}) SMOG survey in the IRAC 3.6--8.0 $\mu$m and MIPS 24 $\mu$m bands. 
We supplement the \emph{Spitzer} data with the data from other satellite and ground-based surveys.
The main goal of our study is to discover and characterize the population of intermediate- and low-mass young stellar objects (YSOs) in the Outer Galaxy and use it to study star formation in a significantly different environment than the Galaxy inside the solar circle. Since the YSOs can be confused with evolved stars in the MIR, these classes of objects need to be carefully separated. Here we present the initial results of our analysis using the K$_{s}$-[8.0] vs. K$_{s}$-[24] CCD as an example. The evolved stars separated from YSOs in the YSO selection process will be investigated in detail in the follow-up study. 
\end{abstract}

\section{Introduction}

\begin{figure}
\begin{center}
\includegraphics[width=\textwidth]{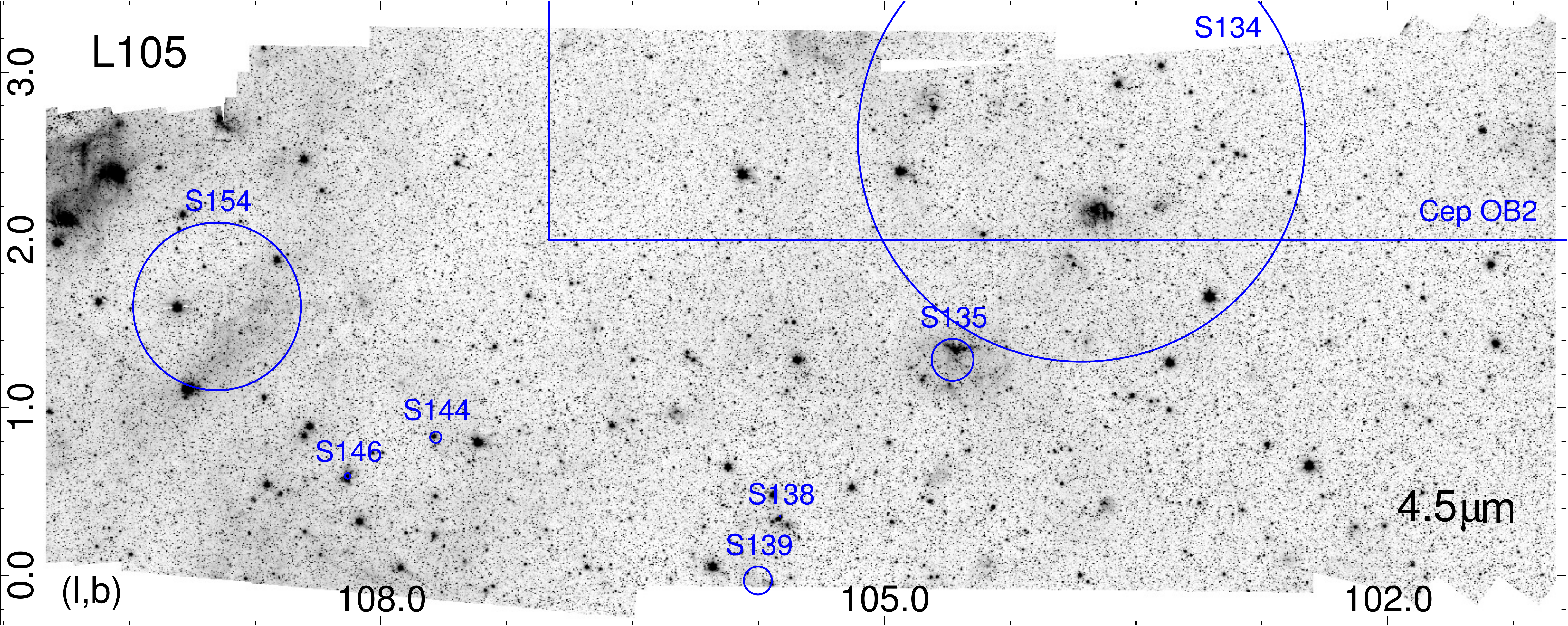}
\end{center}
\caption{\label{Fig1}The IRAC image ($\sim$2$^{"}$ resolution) of L105 region at 4.5\,$\mu$m.
See text for details.}
\end{figure}

Star formation studies in the Galaxy has been concentrating on the nearby molecular clouds (e.g., \cite{Gutermuth_et_al_2008}; \cite{Evans_et_al_2009}; \cite{Megeath_et_al_2012}) or distant massive star formation regions (e.g., \cite{Povich_et_al_2009}), leaving vast clouds of low- and intermediate-mass star formation at larger distances undiscovered. We have initiated a systematic study of star formation in the Outer Galaxy to uncover the population of YSOs in these previously unstudied clouds and investigate the impact of the environment on the star formation process. The \emph{Spitzer} IRAC and MIPS cameras are ideally suited to investigate dusty envelopes of YSOs. We use the data from the ``\emph{Spitzer} Mapping of the Outer Galaxy'' survey (SMOG; PI Sean Carey) that covered $\sim$24 deg$^{2}$ region in the Outer Galaxy: {\it l} = (102$^{\circ}$, 109$^{\circ}$), b = (-0.2$^{\circ}$, 3.2$^{\circ}$) in the IRAC 3.6--8.0 $\mu$m and MIPS 24 $\mu$m bands (see Figure\,\ref{Fig1}). This region that we refer to as ``L105'', samples a range of environments and star formation activities. In Figure\,\ref{Fig1} we marked the boundary of the Cep OB2 association with a blue/black (color electronic version/black and white printout). The H II regions from the \cite{Sharpless_et_al_1959} catalog are shown as blue/black circles and identified by their numbers.

In the Outer Galaxy we expect a contamination from evolved stars such as Asymptotic Giant Branch (AGB) stars, post-AGB objects and planetary nebulae (PNe), or more massive ones such as Red SuperGiants (RSGs), or even Yellow HyperGiants (YHGs). Here, as an example, we concentrate on identifying the location of the low- and intermediate-mass evolved stars (AGBs, post-AGBs and PNe) in the K$_{s}$-[8.0] vs. K$_{s}$-[24] color-color space. We are developing a method to disentagle these classes of sources from the sample of YSO candidates. A detailed description of the YSO selection process will be presented in detail in a future paper.

\section{The K$_{s}$-[8.0]  vs. K$_{s}$-[24] color--color diagram}

The quite robust method to separate different classes of objects, especially when distances to them are unknown, are CCDs and a comparison of new objects with already classified sources or with theoretical models. One of such diagrams, the K$_{s}$-[8.0] 
vs. K$_{s}$-[24] color--color diagram, is known to separate the C-rich from O-rich AGB stars (e.g. \cite{Matsuura_et_al_2014}). In addition, theoretical models from the Grid of RSG and AGB ModelS (GRAMS, \cite{Srinivasan_et_al_2011}, \cite
{Sargent_et_al_2011}) are available. 
The GRAMS grid consists of 66 000 O-rich and 12 000 C-rich models that span the range of stellar and dust shell 
parameters expected for  evolved stars in the
Large Magellanic Cloud (LMC). In the Outer Galay we expect metallicities more typical for the LMC or 
even for the Small Magellanic Cloud (SMC) than for the Galaxy Bulge, so the GRAMS models can be safely applied. On the other hand, there 
are also available self-consistent time-dependent hydrodynamic (HD) radiative transfer calculations for gaseous dusty circumstellar 
shells  around C-rich and O-rich stars in the final stages of their AGB/post-AGB evolution (\cite{Steffen_et_al_1998}). These calculations 
are  based on one  particular stellar evolutionary track and single dust size grain (a=0.05 $\mu$m) for its both chemical compositions: 
amorphous carbon (AC: \cite{RM_1991}) and {\it astronomical} silicates (ASil: \cite{DL_1984}).

\begin{figure}[]
\begin{minipage}{18pc}
\includegraphics[width=18pc]{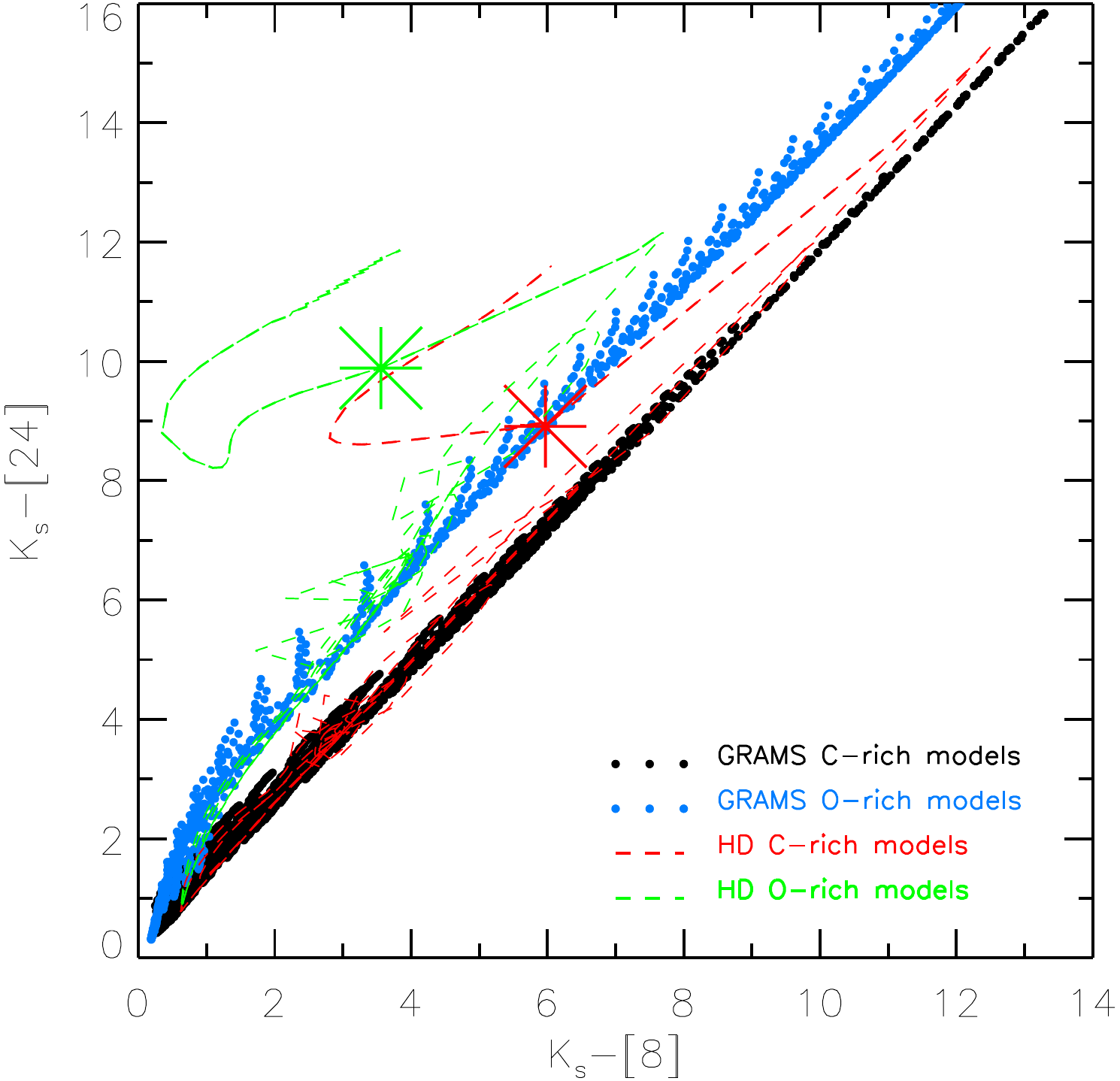}
\caption{\label{Fig2}A comparison between the GRAMS models for RSGs and AGBs from \cite{Srinivasan_et_al_2011} and \cite{Sargent_et_al_2011} with the HD models of AGB and post-AGB phases from \cite{Steffen_et_al_1998} on the  K$_{s}$-[8.0] vs. K$_{s}$-[24] CCD. See the legend and text for details.}
\end{minipage}\hspace{2pc}%
\begin{minipage}{18pc}
\includegraphics[width=18pc]{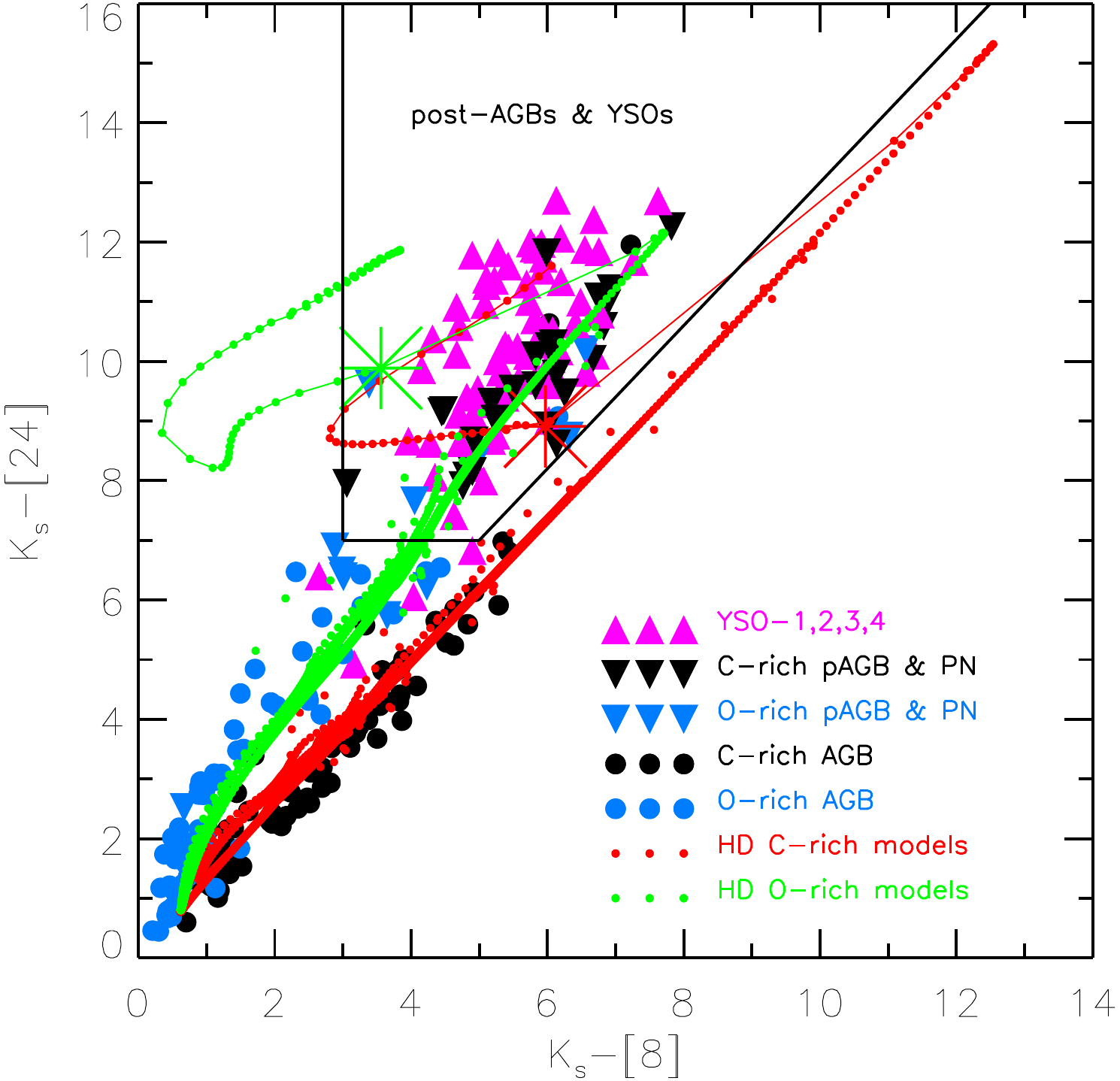}
\caption{\label{Fig3} A comparison between the HD models of AGB and post-AGB evolution from \cite{Steffen_et_al_1998} with sources from the LMC and SMC classified by \cite{Woods_et_al_2011} and \cite{Ruffle_et_al_2015}.  See the legend and text for details.}
\end{minipage} 
\end{figure}

Figure\,\ref{Fig2} shows the K$_{s}$--[8.0]  vs. K$_{s}$--[24] CCD and compares the GRAMS models for C-rich stars (black dots) with the HD
evolutionary track for AC dust (dashed red/dark-gray line), and those for O-rich ones (blue/gray dots) with the HD track for ASil dust (dashed green/light-gray line). The end of the AGB/beginning of the post-AGB phase of evolution (time "0": see \cite{Steffen_et_al_1998} for a definition) is marked 
by the corresponding big star symbol. The large temporal luminosity and mass loss variations, which are associated with thermal pulses during the AGB phase can be recognized as bluer "loops" (see  \cite{Steffen_et_al_1998} for details). Despite the fact that HD models were not adjusted for any specific metallicity, the agreement  between the GRAMS models and the HD evolutionary tracks is quite impressive. The advantage  of the HD models is an additional coverage of the post-AGB phase of stellar evolution (to the left from the big star symbols).   

Figure\,\ref{Fig3} shows the same CCD as Figure\,\ref{Fig2} and compares HD models (same colors as in Figure\,\ref{Fig2}, but know the last 200 years of AGB and about 1000 years of post-AGB evolution are connected by the solid lines) with sources from the LMC and SMC classified by \cite{Woods_et_al_2011} and \cite{Ruffle_et_al_2015}. The C-rich evolved stars are shown by black symbols, while O-rich ones are represented by blue/gray symbols. All types of YSOs defined by \cite{Woods_et_al_2011} and \cite{Ruffle_et_al_2015} based on their spectra are shown by magenta/gray triangles.  The HD models during the AGB phase are plotted in intervals of 100 years (HD models of the post-AGB phase are not equidistant in time), so their number density is a direct measure of the probability of finding objects in different parts of the diagram. Figure\,\ref{Fig3} shows that there is a clear separation between the C-rich and O-rich AGB objects, which can be explained well by the HD models, 
and that post-AGB objects overlap completely with YSOs. The O-rich post-AGB phase of our original HD models seems to be too blue. The reason is too fast drop of flux in [8.0] band in case of {\it astronomical} silicates due to too large optical depth during the transition from AGB to post-AGB phase. This is caused by too small velocity of O-rich dust shell (see Figure\,21 in \cite{Steffen_et_al_1998}).
By black solid lines we have delimited a region of  K$_{s}$--[8.0]  vs. K$_{s}$--[24] CCD, where majority of post-AGB sources and YSOs are located (compare to Figure\,10 of \cite{Matsuura_et_al_2014}). 

\begin{figure}[h]
\includegraphics[width=18pc]{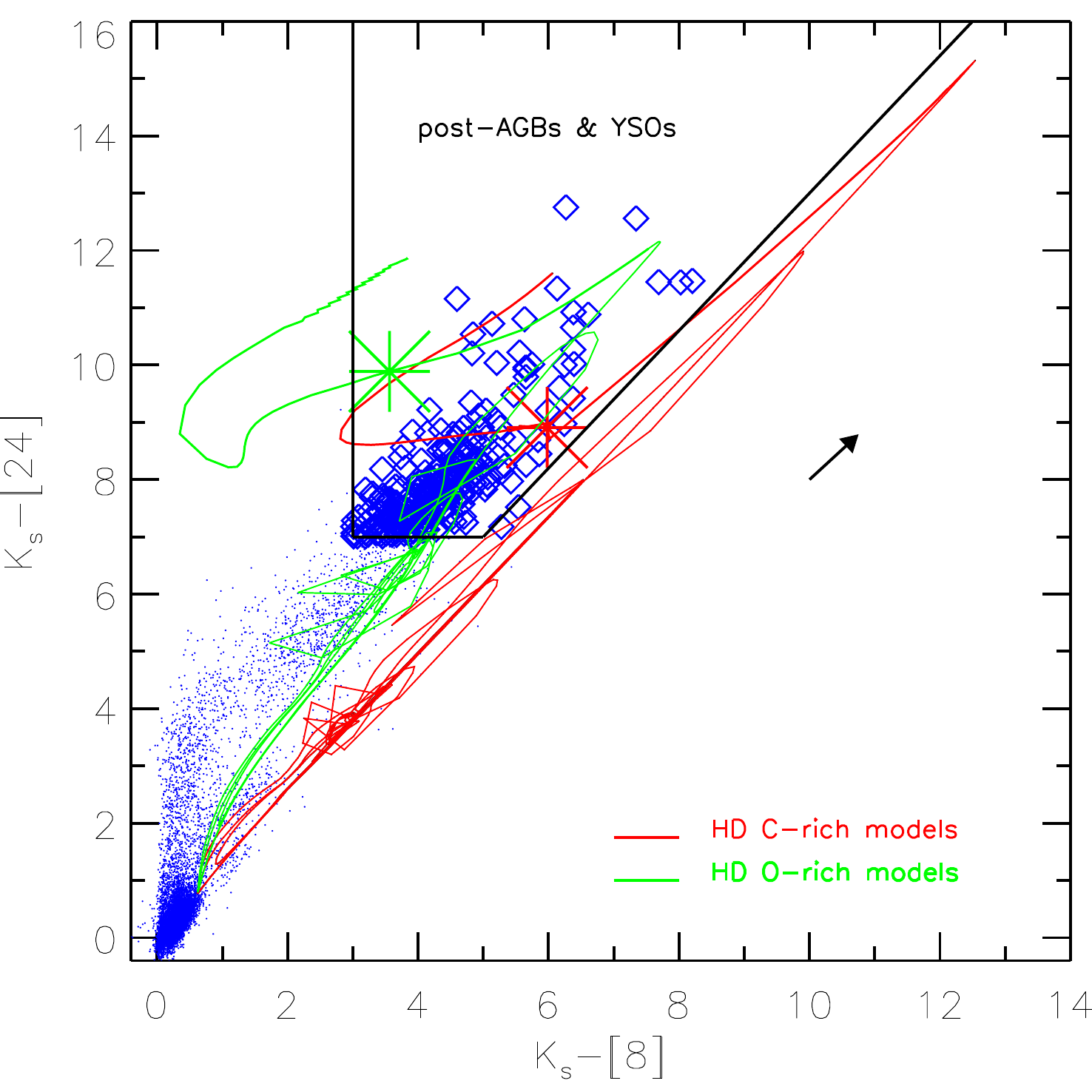}\hspace{2pc}%
\begin{minipage}[b]{18pc}\caption{\label{Fig4}
The K$_{s}$-[8.0]  vs. K$_{s}$-[24] CCD for all SMOG sources (15\,311).
SMOG sources are shown by dots, which are replaced by open large diamonds in the "post-AGBs $\&$ YSOs" region. HD models 
for the AGB and post-AGB phases 
are shown by red/dark gray solid line for C-rich dust and by green/light gray line for O-rich dust. 
The end of the AGB/beginning of post-AGB phase of evolution is marked 
by the corresponding big star symbol. The mass loss at the end of the AGB evolution started to decrease about 200 years before moment "0".
The "post-AGBs $\&$ YSOs" region is marked with the black solid lines (contains 249 objects).
The black arrow shows the reddening vector for A$_V=10$.
}
\end{minipage}
\end{figure}

In Figure\,\ref{Fig4}, the K$_{s}$-[8.0]  vs. K$_{s}$-[24] CCD for all SMOG sources with good quality data at all bands is shown. 
Due to the small number of good quality measurements by MIPS at 24 $\mu$m the total number of plotted objects is only 15\,311. Note, that the total number of sources detected at shorter wavelengths by IRAC is 2\,836\,618. The SMOG sources, which are located in the "post-AGBs $\&$ YSOs" region on this CCD (249) are shown by larger symbols then the rest of the SMOG objects. 
One characteristic result can be seen for SMOG objects located in the L105 Outer Galaxy region: there is relatively small number of C-rich AGB sources located around positions suggested by the HD track for C-rich dust (red/dark gray solid line), and especially those with large mass loss rates (the reddest) are missing. In addition, the reddening vector shown in Fig.\ref{Fig4} constructed for 
A$_V=10$ demonstrates that our HD tracks computed for the lowest mass loss of the order of 10$^{-7}$ M$_{\odot}$\,yr$^{-1}$, will move out of the clump of the SMOG points around K$_S$-[8]$\sim0.8$. 

\section{Summary}

To investigate star formation in the Outer Galaxy we have started analyzing the SMOG data obtained by IRAC and MIPS instruments onboard {\it Spitzer}. We demonstrated that the K$_{s}$-[8.0]  vs. K$_{s}$-[24] CCD separated the C-rich and O-rich AGBs well (see also \cite{Matsuura_et_al_2014}), but this CCD is not able to disentangle YSOs from post-AGB objects. Our color--color cuts in the K$_{s}$-[8.0]  vs. K$_{s}$-[24] CCD select 
$\sim$250 post-AGB objects and YSO candidates. Color-color cuts based on a set of NIR/MIR CCDs will allow us to select an initial sample of YSO and evolved star candidates. To classify these sources and remove contaminants from each sample, we will perform the spectral energy distribution fitting and visual inspection the sources' environments using the multiwavelength images.

\ack
Authors would like to acknowledge financial support from the Polish National Science Center grant  2014/15/B/ST9/02111. R.Sz. and N.S. acknowledge support from the EU FP7- PEOPLE-2010-IRSES program in the framework of project POSTAGBinGALAXIES (Grant Agreement No. 269193) and NCN grant 2011/01/B/ST9/02031. A.K. acknowledges support from the Foundation for Polish Science (FNP) and the Polish National Science Center grant 2013/11/N/ST9/00400. We thank Sean Carey for providing us with the MIPS 24 $\mu$m catalog.

\end{document}